\documentclass[runningheads]{llncs}
\usepackage[T1]{fontenc}
\usepackage{graphicx}
\usepackage{todonotes}
\usepackage{algorithm}
\usepackage{algpseudocode}
\usepackage{amssymb,amsmath}
\usepackage{amsfonts}
\usepackage{caption}
\usepackage{subcaption}
\usepackage{multirow}
\usepackage{microtype}

\usepackage[font=footnotesize]{caption}
\newcommand{\pred}[1]{\text{{#1}}}

\newcommand{\charge}[1]{\pred{HasCharge{#1}}}

\newcommand{\bond}{\pred{Bond}}
\newcommand{\buildingBlock}{\pred{BuildingBlock}}
\newcommand{\amideBond}{\pred{AmideBond}}
\newcommand{\amino}{\pred{AminoResidue}}
\newcommand{\carboxy}{\pred{CarboxyResidue}}

\newcommand{\aar}{\pred{AAR}}
\newcommand{\peptide}{\pred{Peptide}}

\newcommand{\overlap}{\pred{Overlap}}
\newcommand{\atom}{\pred{Atom}}

\begin{document}
\title{ChemLog: Making MSOL Viable for Ontological Classification and Learning}
%
%
\author{Simon Flügel\inst{1}\orcidID{0000-0003-3754-9016} \and
Martin Glauer\inst{2}\orcidID{0000-0001-6772-1943} \and
Till Mossakowski\inst{1}\orcidID{0000-0002-8938-5204} \and
Fabian Neuhaus\inst{2}\orcidID{0000-0002-1058-3102}
}
\authorrunning{S. Flügel et al.}
%
\institute{Institute for Computer Science, University of Osnabr\"uck, Neuer Graben 29, 49074 Osnabr\"uck, Germany \and
Institute for Cooperating Systems, Otto von Guericke University Magdeburg, Universit\"atsplatz 2, 39106 Magdeburg, Germany}
\maketitle              
\begin{abstract}
Despite its prevalence, in many domains, OWL is not expressive enough to define ontology classes.
In this paper, we present an approach that allows to use monadic second-order formalisations for ontology classification.
As a case study, we have applied our approach to 14 peptide-related classes from the chemistry ontology ChEBI. 
For these classes, a monadic second-order logic formalisation has been developed and applied both to ChEBI as well as to 119 million molecules from the chemistry database PubChem.
While this logical approach alone is limited to classification for the specified classes (in our case, (sub)classes of peptides), transformer deep learning models scale classification to the whole of the ChEBI ontology.
We show that when using the classifications obtained by the logical approach as training data, the performance of the deep learning models can be significantly enhanced.

\end{abstract}
%
%


\section{Introduction}
%

The ontological classification of chemical molecules is important for the prediction of chemical properties of molecules, as well as for the discovery of new chemical substances enjoying certain properties.
Ontologies of structured objects (such as chemical molecules) can be used for the classification of new unseen objects (and thus, for the extension of the ontology) in two ways. (1) The structure of the objects can be used to train deep learning models \cite{glauer2024chebifier}. These models directly work on a graph-like structure (e.g., Graph Neural Networks), or on a string representation of structured objects (e.g., Transformers). (2) Alternatively, logical axioms can be used for axiomatising classes of structured objects, the latter being construed as first-order models \cite{kutz2012modelling}. Note that generally, these logical axioms are not part of the OWL formalisation of the ontology, but are specifications of single classes. The reason is that they work on a different level: the domain of discourse is no longer the set of all molecules in the ontology, but rather the set of all atoms in a molecule. These specifications are expressed in a richer language like first-order logic (FOL) or monadic second order logic (MSOL). MSOL has been termed the ``finite automata for graphs'', and (unlike FOL) it can be used to speak about paths in a graph and subgraphs of a graph~\cite{DBLP:books/daglib/0030804}.
Many important chemical classes like peptides are defined through the presences of certain subgraphs and hence cannot be axiomatised in FOL, leave alone in in OWL.
When using MSOL, a key challenge is its model checking complexity and scalability to large datasets.

In this work, we first follow the logical approach and present \textit{ChemLog}, a methodology that can bridge the gap between complex second-order definitions of classes of structured entities and their application to large datasets.
Our approach consists of three stages: In the first stage, we use monadic second-order logic (MSOL) to fully formalise the ontology classes of interest.
In the second stage, we translate these definitions to a set of partial definitions in FOL.
Finally, we turn the FOL definitions into a purely algorithmic implementation.
This allows us to check efficiently which instances fulfil which definitions, while simultaneously ensuring the alignment of the procedural implementation with the original declarative MSOL definitions, thereby enhancing the trustworthiness of the implementation.
Each stage comes with a different type of reasoning: While the algorithmic stage provides a direct result, for the FOL stage, we resort to a model checker. 
For MSOL, model finding with the MONA reasoner~\cite{henriksen1995mona} and a conversion to Quantified Boolean Formulas (QBF) are used.

In a case study on the Chemical Entities of Biological Interest (ChEBI)~\cite{hastings2016chebi} ontology, 14 classes related to peptides have been selected. 
This section of ChEBI includes 14,879 molecules (out of 179,115 in the complete ontology as of version 239). 
For these classes, we have formalised the existing textual definitions into MSOL, taking the instances of classes and definitions from IUPAC \cite{IUPACpeptides} into account to resolve ambiguities. 
On this example, we demonstrate the process of transitioning from MSOL to FOL and an algorithmic solution.
We evaluate the alignment of the layers on the ChEBI ontology and use the algorithmic layer to identify peptides across 119 million molecules from PubChem. 
This shows both that the algorithmic layer is a faithful representation of the MSOL layer, and also that the algorithmic layer is able to cover large datasets that not only cover the size of current domain ontologies, but extend them significantly.

In the second part of our paper, we demonstrate how the results of the logic-base classification can improve the quality of deep learning models that classify new molecules in the ChEBI ontology. 
We train Transformer models (ELECTRA~\cite{clark2020electra}) both on the original ontology, as well as on our reclassification of the 14 ChEBI classes according to our formalised definitions. 
The model trained on our reclassified data improves on the ChEBI-trained model by a significant margin (from a micro F1-score of 85\% to 93\%).
Even on the original ChEBI, the new model achieves a recall of 92\% (compared to 81\% for a model trained directly on ChEBI). 
This indicates that our classification stays true to the intentions of the original ChEBI classes, while improving the consistency and completeness of the taxonomy.

The paper starts with an introduction to our peptide axiomatisation in Section~\ref{sec:peptides}, followed by an explanation of our proposed methodology (Section~\ref{sec:methodology}). In Section~\ref{sec:evaluation}, we show the results of our evaluation on ChEBI and on PubChem. Section~\ref{deep-learning} discusses the effect of the logical classification to deep learning classification. This is followed by a discussion of our results in Section~\ref{sec:discussion}. In Section~\ref{sec:related-work}, we give an overview over related work. Finally, we draw conclusions and give an outlook on future work in Section~\ref{sec:conclusion}.

\section{Axiomatising Peptides}~\label{sec:peptides}
In order to showcase our approach, we will use the concept of peptides in the ChEBI ontology.
%
%
We have developed MSOL definitions for 14 ChEBI classes related to peptides. These classes have been selected because peptides play a central role in ChEBI (about 7.4\% of ChEBI classes are subclasses of peptide). 
However, many molecules in this section of ChEBI are underspecified. Often, they belong to a specific subclass of peptide, e.g. tripeptide, but are only classified as peptides.
Peptides, as defined in ChEBI, are ``Amides derived from two or more amino carboxylic acid molecules (the same or different) by formation of a covalent bond from the carbonyl carbon of one to the nitrogen atom of another with formal loss of water. [...]''.
Thus, our main task for deciding if a molecule is a peptide or not, is to identify the amino acids from which it has been derived.
Regarding amino acids, ChEBI states that ``The term [peptide] is usually applied to structures formed from $alpha$-amino acids, but it includes those derived from any amino carboxylic acid.''
\begin{figure}[tb]{}
     \centering
     \begin{subfigure}[b]{0.28\textwidth}
         \centering
        \includegraphics[width=\textwidth]{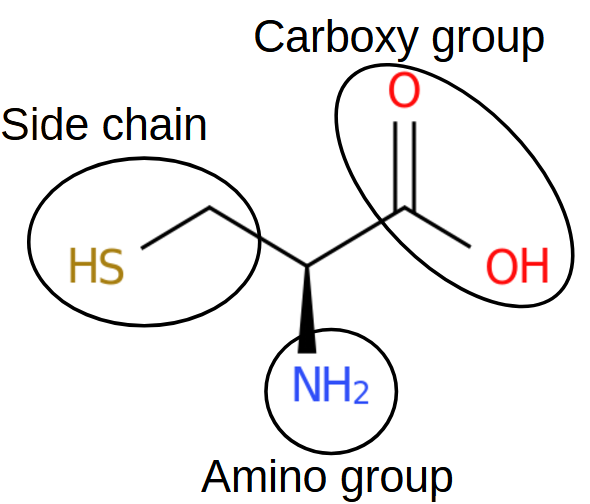}
     \end{subfigure}
     \hfill
     \begin{subfigure}[b]{0.7\textwidth}
        \centering
        \includegraphics[width=\textwidth]{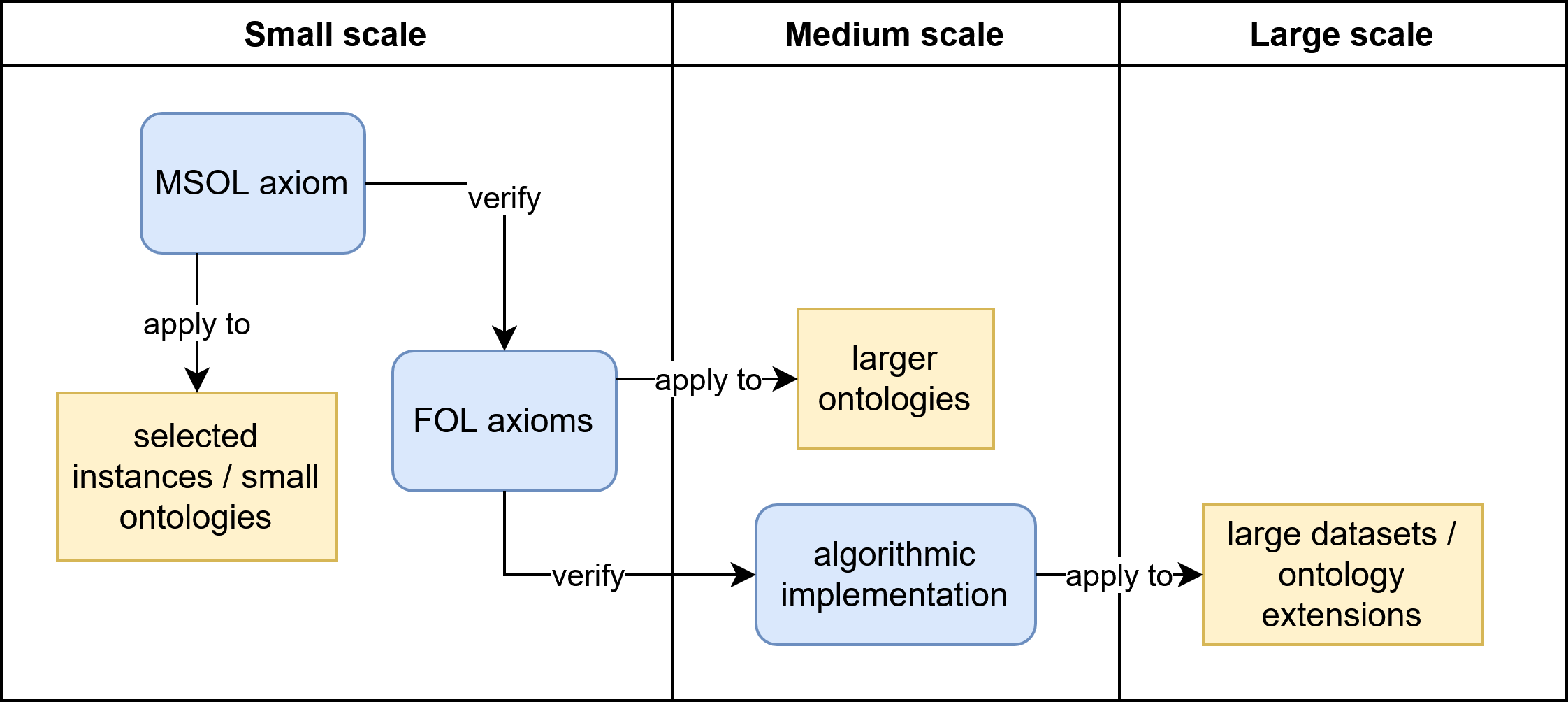}
     \end{subfigure}
     \caption{Left: Structure of the $alpha$-amino acid \textit{L-cysteine}. All $\alpha$-amino acids follow this structure, only with changing side chains. Right: Workflow of our methodology. The MSOL axioms are used to verify their correspondences in FOL, which in turn are used to verify the algorithmic implementation. At the same time, each layer can be used on a different scale.}
     \label{fig:cysteine-plus-architecture}
\end{figure}
Amino acids are defined by the presence of two functional groups, amino groups (predicate name: $\amino$) and carboxylic acids ($\carboxy$). 
In $alpha$-amino acids, they are directly connected by a carbon atom (the amino group is located at the $alpha$ position relative to the carboxylic acid, cf. Figure~\ref{fig:cysteine-plus-architecture}, left).
Since other amino acids are explicitly included, we consider every substructure of a molecule in which the residues of a carboxylic acid and an amino group are connected via a chain of carbon atoms an amino acid residue. 
The predicate $\buildingBlock$ is used to describe fragments of a molecule that are connected via a chain of carbon atoms.
Formally, amino acid residues ($\aar$) are defined in MSOL as follows:
\begin{small}
\begin{equation}
\begin{split}
     \aar(X) \leftrightarrow &(\buildingBlock(X) \wedge \exists a, c_1, c_2, c_3: (\amino(a) \\
     &\wedge \carboxy(c_1, c_2, c_3) \wedge a \in X \wedge c_1 \in X \wedge c_2 \in X \wedge c_3 \in X))
     \label{eq:aar-msol}
\end{split}
\end{equation}    
\end{small} 
For a set of carbon atoms $C$, the notion of a connected subset can be expressed as:
\begin{small}
\begin{equation}\begin{split}
    \pred{CarbonConnected}(X) \leftrightarrow (X \subseteq C \wedge \forall A, B: &(A \neq \emptyset \wedge B \neq \emptyset \wedge A \neq B \wedge A \cup B = X) \\ 
    &\rightarrow \exists u \in A, v \in B: \bond(u,v))
\end{split}\end{equation}
\end{small} 
Note that this predicate requires forming the transitive closure of the bond relation, which cannot be expressed in FOL. 
Therefore, we have to use a more expressive logic, MSOL, to axiomatise peptides.

The final definition of peptides uses a 0-ary predicate ($\peptide$). 
For a peptide, two distinct (i.e., not overlapping) amino acid residues have to be connected by an amide bond ($\amideBond$).
\begin{small}
\begin{equation}\begin{split}
    \peptide \leftrightarrow &(\charge{Neutral} \wedge \exists A_1, A_2, b_1, b_2, b_3: (\aar(A_1) \wedge \aar(A_2)  \\
    &\wedge \neg \overlap(A_1, A_2) \wedge \amideBond(b_1, b_2, b_3) \wedge b_1 \in A_1 \wedge b_3 \in A_2)) 
\end{split}\end{equation}
\end{small}
%
%
%
\label{sec:mol-to-fol-structure}
\begin{figure}[tb]
    \centering
    \includegraphics[width=0.75\linewidth]{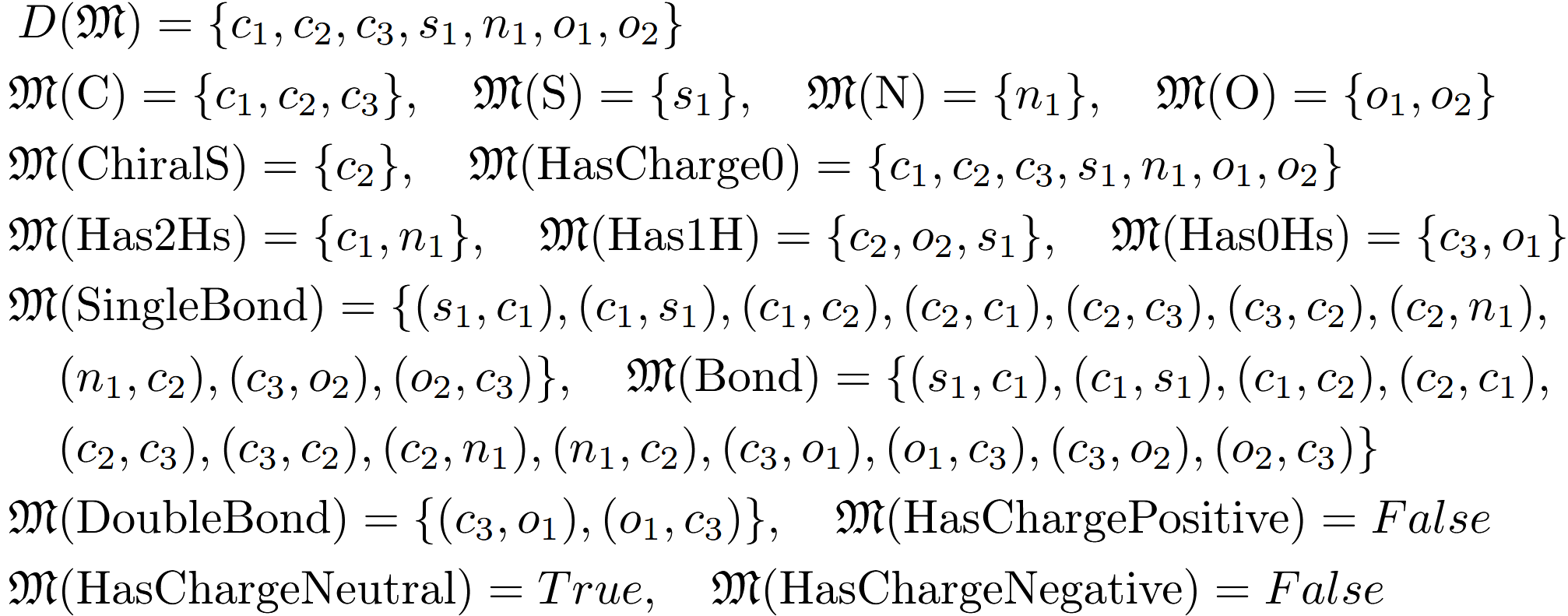}
    \caption{FOL structure for the amino acid \textit{L-cysteine} (cf. Figure~\ref{fig:cysteine-plus-architecture}).}
    \label{fig:fol-structure-cysteine}
\end{figure}
We represent molecules as FOL structures. Figure~\ref{fig:fol-structure-cysteine} shows an example for \textit{L-cysteine}. 
The domain consists of the molecule's non-hydrogen atoms. 
Relevant properties are encoded as unary predicates, bonds between atoms are represented via binary predicates. 
The global charge of the molecule is given by 0-ary predicates.

\section{Logical Methodology}\label{sec:methodology}
This Section describes how the formal definitions of chemical classes are used as a classification tool that determines if a given molecule is part of a class or not. 
For this purpose, we have developed the \textit{ChemLog} framework.
Figure~\ref{fig:cysteine-plus-architecture} (right) gives an overview of the framework's workflow.
This workflow consists of three stages. 
In the first stage, MSOL definitions are used directly. 
For the second stage, a FOL axiomatisation is developed which can be verified by the MSOL axiomatisation.
The third stage is an algorithmic implementation of the MSOL axiomatisation which can in turn be verified by the FOL axiomatisation.

The central idea behind this methodology is that the first stage is a direct implementation of the MSOL definitions. 
However, it is not performant enough to be used on the scale of, e.g., ChEBI with more than 100,000 molecules (see Section~\ref{sec:evaluation}).
Therefore, the second and third stages are iterative steps towards a faster classification and with that classification of larger datasets. 
Nevertheless, this does not make the original MSOL stage superfluous. 
Instead, it is used to validate the correctness of the FOL axiomatisation and of the algorithmic stage. 
While we do not give a formal proof of equivalence between the stages, this allows us to compare the classification results between stages 1, 2 and 3, making the translation's correctness sufficiently plausible.


\subsection{Reasoning in MSOL}
For reasoning in MSOL directly, we use two approaches. 
The first relies on the MSOL model finder MONA~\cite{henriksen1995mona}.
For a given molecule and chemical class, both the structure of the molecule and the class definition are encoded in MONA's MSOL syntax.
For instance, to check if \textit{L-cysteine} (cf. Figure~\ref{fig:cysteine-plus-architecture}) is an amino acid residue, this includes among others the following lines:
\begin{small}
\begin{verbatim}
    var2 Atom, O, A;
    Atom = {0,1,2,3,4,5,6}; O = {5,6};
    pred AAR(var2 X) = BuildingBlock(X) & ex1 A: ex1 C1: ...  
    AAR(A);
\end{verbatim}
\end{small}
The first line declares names as second-order variables. The second states that the variables $Atom$ and $O$ are equivalent to given sets (i.e., there are seven atoms, and two of them are oxygen atoms). The third line defines $\aar$ as a predicate according to Eq.~\ref{eq:aar-msol}. Intermittent definitions have been omitted for brevity. Finally, $A$ is declared to be an amino acid residue.


The second approach encodes the molecule and class definition into a Quantified Boolean Formula (QBF) satisfiability problem. 
This is possible since the size of the domain is limited to the number of atoms. This allows us to represent extensions of predicates as a conjunction of propositional variables, one for each individual (or for each tuple of individuals, for arity >1). Quantification over first-order variables is represented as conjunction or disjunction, and quantification over second-order variables via a series of quantifications over QBF variables.
The above example looks like this in QBF:
\begin{small}
    \begin{equation}\begin{split}
        &\atom_0 \wedge \atom_1 \wedge \atom_2 \wedge \atom_3 \wedge \atom_4 \wedge \atom_5 \wedge \atom_6 \wedge \neg O_0 \wedge \neg O_1 \wedge \neg O_2 \\
        &\wedge \neg O_3 \wedge \neg O_4 \wedge O_5 \wedge O_6 \wedge \exists x_0, x_1, x_2, x_3, x_4, x_5, x_6: (\ldots \wedge \bigvee_{a=0}^6 \bigvee_{c0=0}^6 \bigvee_{c1=0}^6 \bigvee_{c2=0}^6 (\ldots))
    \end{split}\end{equation}
\end{small}
$Atom_i$ states that the $i$-th element in the domain is an atom. $O_i$ states that atom $i$ is an oxygen atom. The last term of the disjunction encodes the definition of amino acid residues (Eq.~\ref{eq:aar-msol}). $x_i$ encodes the presence or absence of atom $i$ in the second-order variable $X$. $\exists a$ is represented with the disjunction $\bigvee_{a=0}^6$.
The resulting QBF problems are converted to CNF using a Tseitin transformation~\cite{tseitin1973reduced}, preprocessed with Bloqqer~\cite{biere2011blocked} and solved by DepQBF~\cite{lonsing2017depqbf}.
In our implementation, MONA and QBF problems (in the QDIMACS format) are automatically generated from an abstract MSOL representation.

\subsection{From MSOL Definitions to FOL Definitions}~\label{sec:meth-msol-fol}
Amino acids can have an arbitrarily long chain of carbon atoms connecting the amino group and the carboxylic acid (cf. Section~\ref{sec:peptides}).
Such a chain cannot be formalised in FOL.
Therefore, when defining amino acids, we can only make statements about amino acids in terms of the carbon-connected ``building blocks'', not define the connection relation itself.
In contrast to MSOL, in FOL the $\buildingBlock$ (cf. Eq.~\ref{eq:aar-msol}) is not logically defined. Instead, it is a primitive predicate whose extension is computed algorithmically and added to a FOL model (see Section~\ref{sec:msol-to-fol-model-checking}). 
The translation from MSOL to FOL is performed automatically in ChemLog. The only predicate that is not automatically translated is $\buildingBlock$, the extension of which is calculated by a manually implemented Python function.
%
%
%
\subsection{From MSOL Model Finding to FOL Model Checking}\label{sec:msol-to-fol-model-checking}
For FOL, model checkers are available and easily implemented. Hence we can directly use the original problem formulation and check whether the FOL axiomatisation of a given class holds in a FOL structure that represents a molecule (like the example in Figure~\ref{fig:fol-structure-cysteine}). 
%
As discussed in Section~\ref{sec:meth-msol-fol}, the building blocks for amino acids have to be calculated algorithmically. Thus, the structure from Figure~\ref{fig:fol-structure-cysteine} is augmented with the carbon-connected building blocks. 
The building blocks are connected to the atoms they include via a binary predicate.
To verify that the structure's augmented part, i.e., the extension of $\buildingBlock$, corresponds to the MSOL axiomatisation, MONA model finding can be used either directly on the elements of the extension or for the final classification.

\subsection{Model Checking in FOL}


We have implemented a custom FOL model checking algorithm in Python. 
The algorithm expects formulas in prenex normal form and a matrix in conjunctive normal form.
It applies the following steps recursively for a given sentence $\phi$ and structure $\mathfrak{M}$:
\begin{enumerate}
    \item Simplify the sentence by removing all literals that contain no variables and are false. If this results in an empty clause, the algorithm terminates as no model of the formula exists. 
    \item Remove all clauses without variables. Since false literals have been removed, any remaining non-empty clauses without variables must be true.
    \item Find all clauses that contain exactly one variable. Then, for each variable $v$ and element in the domain $u$, apply the model checking algorithm to $\mathfrak{M}[c \mapsto u]$ and the individual clauses of $\phi[v \mapsto c]$ containing the constant $c$. 
    \item Sort the variables by the number of domain elements which can be used to replace them. If there are none, the algorithm terminates. Otherwise, apply the algorithm recursively to $\mathfrak{M}[c \mapsto u]$ and $\phi[v \mapsto c]$, now using the complete formula instead of only selected clauses.
\end{enumerate}

The idea behind the last two steps is to select the variables that have the lowest number of candidates first. In pratice, this leads to linear instead of exponential (in the number of variables) complexity for chemical structures: the model checker will start with heteroatoms and then move along bonds.


\subsection{From FOL to an Algorithmic Implementation}
While FOL model checking is more efficient than MSOL theorem proving, it might still be insufficient for large datasets.
Therefore, we have implemented an automatic translation from FOL to Python. Each predicate corresponds to a Python function which accepts an RDKit~\footnote{\url{https://www.rdkit.org/}} molecule object (\verb|mol|), the second-order objects that have already been calculated algorithmically (i.e., the extension of $\buildingBlock$, cf. Section~\ref{sec:meth-msol-fol}, \verb|so_elements|) and the arguments of the FOL predicate. For example, for amino acid residues, we get the following Python function:
\begin{small}
    \begin{verbatim}
def AAR(mol: Chem.Mol, so_elements, x):
    return any((x in so_elements and AminoResidue(mol, so_elements, x0) 
        and x0.GetIdx() in x and CarboxyResidue(mol, so_elements, x1, x2,
        x3) and x1.GetIdx() in x) for x0 in mol.GetAtoms() for x1 in 
        mol.GetAtoms() for x2 in mol.GetAtoms() for x3 in mol.GetAtoms())
    \end{verbatim}  
\end{small}
The final algorithmic implementation has been further optimised for efficiency. 
Overall, this gives us three different strategies (including two different implementations for the MSOL strategy) for molecule classification: Fully reasoning-based, partially algorithmic and fully algorithmic.
The purpose of this is to provide validation tools (one strategy or implementation can validate another) and to provide scalability, as different strategies provide access to different reasoning tools or, in the case of the algorithmic implementation, do not rely on reasoning tools at all.
The implementation for ChemLog is available on GitHub~\footnote{\url{https://github.com/sfluegel05/chemlog-peptides}}. A web front-end for molecule classification can also be accessed as part of Chebifier~\cite{glauer2024chebifier}~\footnote{\url{https://chebifier.hastingslab.org/}. Select ``ChemLog Peptides'' to get only results from ChemLog as discussed in this work.}

\section{Evaluation of Logical Classification}~\label{sec:evaluation}
We apply our approach to increasingly large datasets, starting with the smallest 1,000 peptides from ChEBI, on which we verify the alignment of the MSOL and FOL formalisations. Here, we focus on the classification by number of amino acids (i.e., into dipeptide, tripeptide, ...).
Then we move on to the complete 3-STAR (= high-quality) subset of ChEBI with 45 thousand molecules. 
There, we verify the alignment of the FOL and algorithmic approaches. 
Besides the size-based classification, this also includes charge-based (peptide anion, zwitterion, ...) and proteinogenic amino acid (glycine, alanine, ...) classification.
Finally, we apply the algorithmic approach to 119 million molecules from PubChem.

\subsection{Layer Alignment}
We have compared the three layers according to two criteria: Do all layers yield the same results? And how fast is each approach?
In a first step,   the comparison was based on the 1,000 shortest molecules that are classified as peptides in ChEBI, where length is calculated based on   SMILES strings in ChEBI \cite{weininger1988smiles}. 
\begin{table}[tb]
    \centering
    \begin{footnotesize}
    \begin{tabular}{|c|c|c|c|l|}
        \hline
       Dataset~~~ & Layer & Failure rate & Different to MSOL/FOL & Time/molecule (s) \\
       \hline
       & MSOL-MONA & 61.1\% & - & 10.8 \\
       & MSOL-QBF & 0 & 0 & 832 \\
       1,000 & FOL & 0.4\% & 0 & 0.37 \\
       & Algorithmic & 0 & 0 & 0.0015 \\
       \hline
        \multirow{2}{3.5em}{\mbox{3{-}STAR} 45,000} & FOL & 2.2\% & - & 1.03 \\
       & Algorithmic & 0 & 0 & 0.0084 \\
       \hline
    \end{tabular}
    \end{footnotesize}
        \caption{Results of our evaluation on the smallest 1,000 peptides and on the 3-STAR subset of ChEBI. For each layer, the table shows the number of failed proof attempts, number of results that differ compared to MSOL (for the 3-STAR  subset, compared to FOL) and mean time per molecule.}\label{tab:eval-1000-peptides}
\end{table}

For this dataset, all 4 methods gave the same results (excluding failures). 
The failure rates and times per molecule for this evaluation are shown in Table~\ref{tab:eval-1000-peptides}.
The MONA fails to classify more than half of the molecules. 
This is caused by non-configurable constraints of the MONA reasoner, which terminates if the internal graph reaches a certain size or a memory limit is exceeded.
For the FOL model checker, we have applied a 30 second timeout, leading to failures for 4 of the molecules.

Overall, the classification with QBF is the slowest and the algorithmic layer is the fastest, with a difference of 5 orders of magnitude. 
QBF is slowest since the conversion of the problem to QBF leads to large problem sizes. For instance, to identify if a molecule with 16 atoms is a peptide, 136,000 variables and 539,000 clauses are generated. 
Despite the problem size, with specialised preprocessors, we are still able to reliably get results without running into memory limitations (contrary to MONA).

In comparison to FOL, MONA is significantly slower because it has to solve a model finding problem, instead of a model checking problem.
In FOL, the molecule representation is not given axiomatically, but as a FOL structure, see Section~\ref{sec:msol-to-fol-model-checking}.
MONA is also not optimized for the given task. 
The Python implementation uses several graph algorithms that are specialised for tasks such as determining the connected subgraphs of a graph. 
The FOL implementation uses some of these algorithms at critical points as well (mainly, the identification of building blocks for finding amino acids).
Contrary to MONA, the FOL model checker uses a heuristic that is specifically tailored for chemistry-related tasks. 


To get a broader picture of the performance of the FOL and algorithmic layer on larger molecules, we ran them on all molecules from the 3-STAR subset of ChEBI (45,450 instances). 
For all molecules where both implementations returned a classification, the classification was the same.
This is the result of our development process in which ChEBI molecules have been used to debug the algorithmic implementation and to verify its correctness with regard to the FOL axiomatisation.
The same goes for the alignment between the MSOL and FOL layers.
%
The FOL implementation failed to classify some of the larger molecules either due to memory limitations or after reaching a timeout.
Time-wise, both the FOL and algorithmic implementations were slower for this dataset compared to the 1,000 smallest peptides.
This is due to larger molecule sizes in this dataset.
\subsection{Classifying PubChem Data}
We have classified molecules from PubChem~\cite{kim2025pubchem}, a public chemical database covering 119.3 million compounds.
PubChem integrates data from over 1,000 different sources, covering a wide array of domains, including safety, biological targets or drug discovery. 
It cross-references to scientific literature as well as other databases such as ChEBI.
Due to its size, building a taxonomy similar to that of ChEBI manually is not feasible for PubChem. 
However, we can use our automated approach to sort PubChem compounds into our defined classes.

\begin{figure}[tb]
    \centering
    \includegraphics[width=0.7\linewidth]{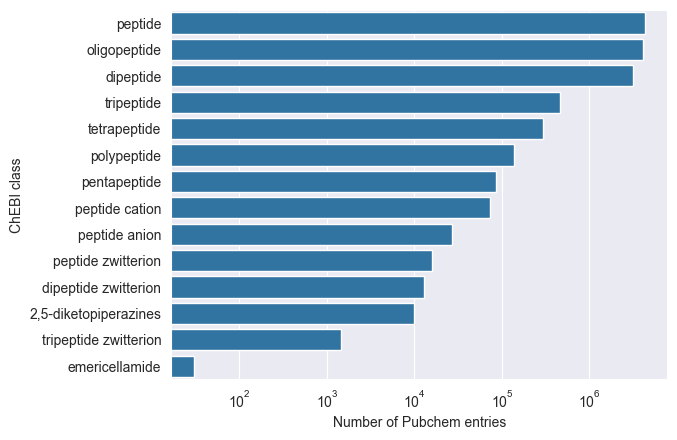}
    \caption{Number of PubChem compounds assigned to ChEBI classes.}
        \label{fig:pubchem-classwise}
\end{figure}
We used the compounds in SDF format provided by the PubChem FTP-server\footnote{\url{https://ftp.ncbi.nlm.nih.gov/pubchem/Compound/CURRENT-Full/SDF/}}. 
Our classification failed for 70 molecules due to reaching a size limit at 10,000 carbon connected fragments. 
In total, we found that 4,419,232 molecules (3.7\% of PubChem) belong to at least one of our classes, the vast majority of which are dipeptides (cf. Figure~\ref{fig:pubchem-classwise}). 
The classification took 2 days and 9 hours, which corresponds to $0.0017$ seconds per molecule. 
This is more than on the ChEBI subsets, most likely due to the broader scope of PubChem, which also includes larger molecules, while ChEBI mostly focuses on small molecules.


\section{Enhancing Performance of Deep Learning Models}\label{deep-learning}
Recall that our task is the classification of (new) molecules into ChEBI classes. Since these classes can overlap, this is a multi-label classification problem.
While ChemLog performs well on peptides, this covers only a fraction of ChEBI. Classification based on the ChemLog methodology requires the design of a large amount of logical axioms, which currenly does not scale to the whole of ChEBI. Therefore, we also have to rely on machine learning models. That said, we here consider the restriction to classification into peptide classes, keeping in mind that an extension to all of ChEBI is much more easily possible for machine learning than for ChemLog.
We interpret this classification as a multi-label machine learning task, using SMILES, a string representation of molecule graphs, as inputs. 
To construct the data set, all SMILES-annotated ChEBI classes were taken as instances and the 14 peptide-related classes were used as labels. 
We are interested in comparing two different ways of assigning labels to instances: The first uses the subsumption relations provided by ChEBI, the second uses our ChemLog methodology.
\begin{table}[tb]
    \centering
    \caption{Number of instances in different classes using either ChEBI subsumption axioms or the ChemLog classification.}
    \begin{footnotesize}
    \begin{tabular}{|c|c|c|c|}
    \hline
        Label class & \#Instances & \#Instances & \#Chemlog / \\
        & ChEBI & ChemLog & \#ChEBI \\
    \hline
    peptide & 14767 & 18137 & 1.23 \\
peptide zwitterion & 91 & 142 & 1.56 \\
dipeptide zwitterion & 66 & 109 & 1.65 \\
emericellamide & 6 & 10 & 1.67 \\
tripeptide zwitterion & 8 & 15 & 1.88 \\
polypeptide & 470 & 1018 & 2.17 \\
peptide anion & 105 & 303 & 2.89 \\
peptide cation & 46 & 171 & 3.72 \\
oligopeptide & 4510 & 17112 & 3.79 \\
dipeptide & 1216 & 4795 & 3.94 \\
2,5-diketopiperazines & 69 & 311 & 4.51 \\
tetrapeptide & 134 & 704 & 5.25 \\
pentapeptide & 21 & 719 & 34.24 \\
tripeptide & 221 & 9077 & 41.07 \\
\hline
    \end{tabular}
    \end{footnotesize}
    \label{tab:instances-per-class}
\end{table}
Table~\ref{tab:instances-per-class} shows the number of instances for both datasets. 
The distribution between classes is roughly similar to that of PubChem (cf. Figure~\ref{fig:pubchem-classwise}), with \textit{peptide} being the largest and \textit{emericellamide} being the smallest class. 
Comparing ChEBI and ChemLog labels, ChemLog assigns more instances to each class overall, increasing the class size by 23\% for peptides up to a factor of 41 for tripeptides. 

This can be explained by an incompleteness in the ChEBI ontology. 
ChEBI only makes positive statements about subclass relations, but no negative ones. 
Thus, many molecules are not assigned to all classes they could be assigned to.
While, for example, \textit{vancomycin} (CHEBI:28001) is undeniably an oligopeptide, it being a glycopeptide and its role as an antibacterial drug have most likely been deemed more important by the curators. 
Thus, no oligopeptide annotation has been made.
The most extreme example for this is the classification of tripeptides, which are easy to recognise by their names put together out of 3 amino acids, e.g., \textit{Val-Tyr-Trp}, but which are only classified as peptides or oligopeptides.
As of March 2025, this issue has been addressed in ChEBI. 
Our evaluation however is based on version 239 of ChEBI, published in January 2025.

\begin{table}[tb]
\begin{center}
\begin{footnotesize}
\begin{tabular}{ |c|c|c|c|c|c|c|c| } 
\hline
Trained on & Tested on & Micro-F1 & Micro-Precision & Micro-Recall \\
 \hline
ChEBI & ChEBI & $0.847 \pm 0.004$ & $0.893 \pm 0.020$ & $0.806 \pm 0.012$ \\
ChEBI & ChemLog & $0.512 \pm 0.011$ & $\mathbf{0.937 \pm 0.010}$ & $0.352 \pm 0.011$ \\
ChemLog & ChEBI & $0.552 \pm 0.005$ & $0.394 \pm 0.007$ & $\mathbf{0.920 \pm 0.008}$ \\
ChemLog & ChemLog & $\mathbf{0.932 \pm 0.002}$ & $\mathbf{0.945 \pm 0.014}$ & $\mathbf{0.919 \pm 0.009}$ \\
 \hline
\end{tabular}
\end{footnotesize}
\end{center}
\caption{Average global metrics and standard deviations for different combinations of training and test sets.}\label{tbl:nn-global-metrics}
\end{table}
The datasets are partitioned using a stratified split with the proportions 85\% / 2.25\% / 12.75\% for the train, validation and test set.
Both datasets use the same split, i.e., the same instances are in the test set for both datasets. 
This allows us to test models trained on one dataset on the other dataset.
We use an ELECTRA model~\cite{clark2020electra} which has been pre-trained on SMILES strings from PubChem.
On each dataset, based on the same pre-trained model, 5 models each have been fine-tuned. 
Table~\ref{tbl:nn-global-metrics} reports the average performance scores and standard deviations.

Overall, the ChemLog models (i.e., models trained with ChemLog-assigned labels), tested on ChemLog, perform best, with an F1-score of 93.2\% compared to 84.7\% for the ChEBI models, tested on ChEBI.
Unsurprisingly, the F1-scores of both models decrease drastically if tested on the other dataset. 
This is to be expected since we are testing the models on a different distribution compared to their training set.
However, that drop in performance takes two different forms. 
When splitting the F1-score into its components, precision and recall, it becomes apparent that the recall does only decrease for the ChEBI models while the precision only decreases for the ChemLog models.
This suggests that the concept of peptides, as learned by the ChemLog models, is not independent to that of the ChEBI models, but rather an extension. 

In addition, the recall on the ChEBI test set is higher for the ChemLog models (92\%) compared to the ChEBI models (81\%). 
This means that, given a peptide in ChEBI, the ChemLog models are more likely to identify it correctly.
Another interesting result is that the precision of ChEBI models is higher tested on ChemLog (93.7\%) compared to tested on ChEBI (89.3\%). 
A likely explanation for this behaviour is that ChemLog provides a ``cleaner'', more consistent version of the ChEBI data. 
\begin{figure}[tb]
    \centering
    \includegraphics[width=1\linewidth]{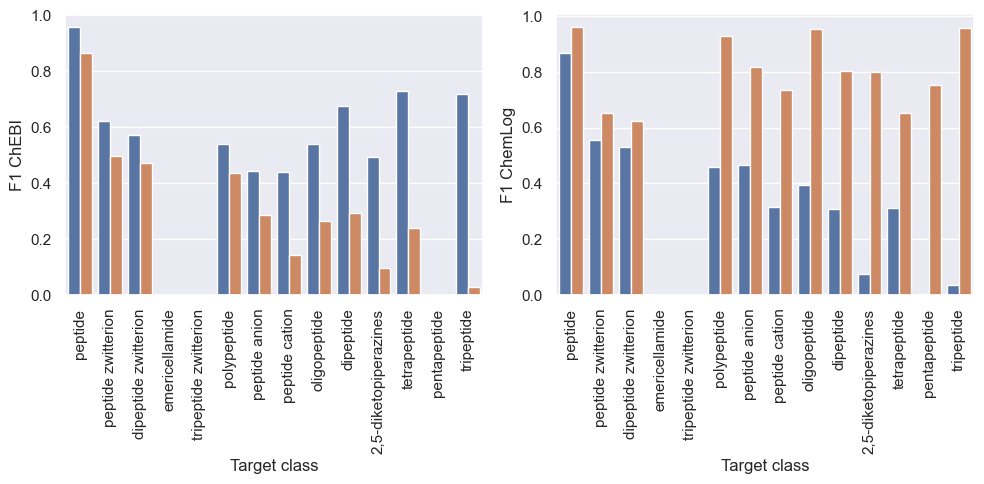}
    \caption{F1-score by label class for ChEBI-trained (left) and ChemLog-trained (right) models. For each class, the performance on the ChEBI (blue) and the ChemLog (orange) test set is reported.}    \label{fig:f1-classwise}
\end{figure}
Figure~\ref{fig:f1-classwise} shows the F1-scores for each target class. 
We can see that the performance differs widely between classes, reaching 96\% on peptides for both ChEBI- and ChemLog-trained models and resting at 0 for tripeptide zwitterions and emericellamides for all models.
This is expected since peptides are a well-populated class with 15,000 / 18,000 instances in both datasets. For emericellamides on the other hand, the models train on only 4 / 8 samples (with 2 left for the test set). 
Given that the train set consists of 159,642 instances, without specialised techniques such a oversampling or weighting, the rare positive samples get drowned out.

For some classes, the performance is roughly the same when comparing ChEBI and ChemLog models, e.g., for \textit{peptide} and \textit{peptide zwitterion}. 
For others, the ChemLog models perform significantly better, e.g., for \textit{polypeptide} and \textit{tripeptide}.
Three classes are noteworthy in particular: \textit{tetrapeptide} is the only class where, on average, ChEBI models outperform the ChemLog models. 
\textit{Peptide anion} is the only class where ChemLog models slightly outperform the ChEBI models on the ChEBI test set. 
Finally, \textit{pentapeptide} is a class that has not been learnable or the ChEBI models, but became learnable for ChemLog models. 
This is due to the larger number of samples available for ChemLog (719 vs. 21 in ChEBI) and the cleaner definition in ChemLog. 

\section{Discussion}~\label{sec:discussion}
Our evaluation has three main findings: 
First, it shows that the ChemLog methodology can ensure that the MSOL axiomatisation is applied correctly. 
We were able to verify on a set of 1,000 peptides that the direct application of the MSOL axiom yields the same results as the FOL and algorithmic stages.
Furthermore, on a large subset of ChEBI with 45 thousand structures, our evaluation has shown that the algorithmic implementation corresponds with the FOL axiomatisation.
While this does not constitute a formal proof of correctness, it is enough to be trustworthy in practical applications.
The main advantage of this test case-based approach is that it is independent of the exact implementation or the target classes. 
Further trustworthiness is achieved by the way that the implementation is derived from the logical axiomatisation: certain predicates are pre-computed or even replaced by Python functions. That is, the implementation follows the logical structure.
This  methodology can be easily expanded to new classes. 
We also propose to use the test cases as a validation tool during the development of the axiomatisation, iteratively refining all three stages. 

Second, we have established that our approach can be applied to large datasets. 
With $0.00047$ seconds per molecule on ChEBI and $0.0017$ seconds on PubChem, the algorithmic stage can be used to classify large amounts of data in a reasonable amount of time. 
Taking into account that many chemical definitions are less complex than the one of peptides and further optimisations are possible (e.g., by improving the implementation for existing classes or using the taxonomy to only look for peptides if a superclass of peptides has been identified), we believe that the ChemLog methodology has the potential to be used broadly in the development of ontologies.

Third, we have shown how, in a machine learning environment, automatic classification can improve performance while still connecting back to the ontology developers' intentions. 
Overall, models trained on ChemLog have an 8 percent higher F1-score compared to models trained on ChEBI.
In particular, some classes which were hard to learn in the original dataset now yield significantly higher scores. 
This is due to two mechanisms: 
(a) The automatic classification fills in the gaps in the ontological classification. 
Often, ontologies do not provide a complete OWL axiomatisation of their classes. 
For instance, in ChEBI, not all molecules that are oligopeptides have an axiom \textit{is an oligopeptide}. 
This is a problem for machine learning models, because they require large amounts of data.
Giving them less samples for a class makes it harder to generalise to unseen members of the class.
Even worse, since the oligopeptides that are not annotated as such still exist in the ontology, a Machine Learning model will learn them as negative samples, i.e., molecules that are not oligopeptides.
This actively hurts their performance.
%
(b) The automatic classification removes errors and inconsistencies from the dataset.
While ChEBI adheres to a high standard of human curation, as any ontology of this size, it is bound to have errors. 
We have spotted some of these errors during our evaluation. For instance, Leu-Gly-Pro (CHEBI:6414) was classified as a dipeptide. 
This error has since been reported and fixed by the ChEBI development team. 
Also, independent curators might interpret definitions differently. 
For example, oligopeptides are defined as ``containing a relative small number of amino acids''. 
Deciding how many amino acids are still considered ``small'' is up to the curator. 
In some cases, e.g. for CHEBI:136715, this can mean up to 19 amino acids. 
In other cases, peptides with 10 amino acids are already considered polypeptides (e.g., \textit{abarelix}, CHEBI:337298).
The process of formalising these definitions lays open such ambiguities and establishes clear rules for what belongs to a class and what does not (in the case of oligopeptides, only molecules with up to 9 amino acids).

When developing a formal axiomatisation of ontology classes, it is important to stay true to the intentions of the ontology developers. 
Otherwise, we would end up with a new class that, while properly axiomatised, has no direct mapping to the ontology class.
We have ensured this alignment by taking into account the textual definitions of ChEBI and IUPAC, as well as the already classified instances. 
In addition, we have collaborated with domain experts to iteratively improve our definitions and cover edge cases.
In our evaluation, the high recall scores of ChemLog models on the ChEBI test set indicate that the alignment was successful. 
The Transformer models, as independent agents with no prior knowledge of peptides, were given the task to infer a general concept of peptides based on examples from the ChemLog classification.
The recall score of 92\% on both the ChemLog and ChEBI test sets shows that the learned concept of peptides matches the ChEBI and ChemLog classifications equally. 

\section{Related Work}~\label{sec:related-work}
Over the last years, different methods have been proposed for automatic ontology extensions in the context of chemistry, include rule-based and Machine Learning-based approaches.
\cite{glauer2024chebifier} introduces a Transformer-based model for classification in ChEBI, covering the 1,332 most populated classes of ChEBI. 
They use SMILES strings provided by ChEBI as their training data and extract labels from the ChEBI taxonomy. 
A weighting scheme is used to counteract the class imbalance problem, leading to improved performance for smaller classes.
In the context of this work, we see Machine Learning as a supplementary tool to rule-based classification. 
As we have demonstrated in Section~\ref{sec:evaluation} for peptides, some classes lack the size needed for Machine Learning (e.g., emericellamides) or are incomplete in ChEBI (e.g., tripeptides), making them not suitable for data-driven models.
Many other classes however are relatively easy to learn or cannot be defined structurally (e.g., peptide antibiotic).
Especially since ML models require no human intervention, we propose to use them as a ``default'' classification tool and to target specific parts of the hierarchy with our approach.

A comprehensive effort for a rule-based chemical classification has been conducted by ClassyFire~\cite{djoumbou2016classyfire}. 
Unfortunately, the ClassyFire's rules are not publicly available. In earlier work, we could outperform ClassyFire using Transformers \cite{glauer2024chebifier}.
An OWL-based approach that allows for a high degree of automation is the self-organising ontology proposed by Chepelev et al.~\cite{chepelev2012self}. 
Based on a set of chemicals manually assigned to a certain class, reoccurring features are automatically detected and combined into an OWL definition. 
This definition is then tested and iteratively improved. 
They apply their approach to 60 MeSH and 40 ChEBI classes, which can then be automatically sorted into a hierarchy. 

In~\cite{flugel2022fowl}, a method for assigning FOL annotations to OWL classes, FOWL, has been developed and applied to the ChEBI ontology. 
The existing SMILES annotations were automatically translated into FOL. The resulting ontology, consisting of FOL and OWL axioms, has been verified by a FOL reasoner. 
With ChemLog, we provide a more general framework than other rule-based approaches for several reasons: For one, we use a more expressive logic.
We have shown that a complete definition of peptides in FOL is not possible. 
The same is true for SMARTS \cite{weininger2012smarts}.
Also, our approach is not limited to chemistry as we use a general logic, MSOL.
Therefore, one could apply ChemLog to other domains as well. 

A formalisation of chemical concept into MSOL has first been developed by us in~\cite{kutz2012modelling} for \textit{fullerenes}, a class of molecules with a complex 3-dimensional structure (which is a niche topic). 
However, in~\cite{kutz2012modelling} we do not provide an implementation of this framework. 
ChemLog bridges this gap between the MSOL formalisation of classes and their practical application to OWL ontologies. 
With the application of our methodology to peptides, we have demonstrated that MSOL formalisations are necessary for a large portion of the chemical domain.

\section{Conclusion}~\label{sec:conclusion}
In this paper we introduced ChemLog, a methodology that harnesses monadic second-order formalisations for ontology classification via FOL model checking and algorithmic implementations.
In a case study on the ChEBI ontology, we have developed MSOL definitions for 14 peptide-related classes. 
For this example we showed how MSOL definitions may be directly translated to FOL (and an extended FOL structure) as well as be implemented algorithmically.
The evaluation on the ChEBI ontology has shown that the three stages of our approach can be used to verify each other and provide an efficient debugging mechanism.
Using PubChem, we have demonstrated that ChemLog can be used to scale MSOL definitions up to datasets with more than 100 million instances.
Finally, we use the results of the logic-based classification to improve the quality of deep learning models.
In future work, we will improve the transition from logically defined predicates to algorithmic definitions, using the experience gathered on peptides. This will include automating the translation of second-order predicates and improving the efficiency of the FOL definitions and Python functions resulting from the translation.
The general methodology employed in this paper does not make use of specific properties of peptides. Hence,
we expect that we can scale up our approach towards other classes in ChEBI as well as ontologies of structured entities in different domains.

\subsubsection{\ackname} 
This work has been funded by the Deutsche Forschungsgesellschaft (DFG, German Research Foundation) - 522907718 and 456666331.

\subsubsection{Disclosure of Interests.}
The authors have no competing interests to declare that are
relevant to the content of this article. 
%
%
%
\bibliographystyle{splncs04}
\bibliography{bibliography}

\end{document}